\documentclass[twocolumn, 10pt, journal]{article}

\usepackage[utf8]{inputenc}
\usepackage[margin=0.8in]{geometry}
\usepackage{authblk}
\usepackage{setspace}
\usepackage{graphicx}
\graphicspath{{./figures/}}
\usepackage{subcaption}
\usepackage{amsmath}
\usepackage{amssymb}
\usepackage{booktabs}
\usepackage{hyperref}
\usepackage{xcolor,paralist,hyperref,titlesec,fancyhdr,etoolbox}
\usepackage[defaultcolor=red]{changes}
\usepackage{multirow}

\title{Multi-Objective Optimization with Quantum-Annealing-Inspired Algorithms}

\author[1,2]{Xian-Zhe~Tao\textsuperscript{$\dag$}}
\author[3]{Pavel~Mosharev\textsuperscript{$\dag$}}
\author[4]{Zhaofeng~Su}
\author[2]{Man-Hong~Yung\textsuperscript{*}}

\affil[1]{Southern University of Science and Technology, Shenzhen, 518055, China}
\affil[2]{International Quantum Academy, Shenzhen, 518048, China}
\affil[3]{School of Physical Sciences, University of Science and Technology of China, Hefei, 230026, China}
\affil[4]{Research Institute of Quantum Technology, Hong Kong Polytechnic University, Hong Kong, China}
\date{\today}

\begin{document}
\onecolumn

\maketitle

\thispagestyle{fancy}       % 仅对当前页（首页）应用 fancy 样式
\fancyhf{}                  % 清空当前的页眉和页脚内容
\renewcommand{\headrulewidth}{0pt} % 确保首页没有页眉线（如果需要保留页眉请注释掉这行）
\renewcommand{\footrulewidth}{0.4pt} % 添加页脚分割线 (可选，不喜欢可设为 0pt)
% 设置页脚内容 (居中显示，也可以改为 [L] 左对齐 或 [R] 右对齐)
\fancyfoot[L]{%
    \begin{minipage}{0.9\textwidth}
        % \centering
        \footnotesize
        \noindent
        \textsuperscript{ * }Address correspondence to: yung@iqasz.cn \\
        \vspace{0pt} % 调整两行之间的间距
        \textsuperscript{$\dag$}These authors contributed equally to this work.
    \end{minipage}
    }
% \let\thefootnote\relax
% \footnotetext{MSC2020: Primary 00A05, Secondary 00A66.} %%%%%%%%%%

\begin{abstract}

Combinatorial optimization is widely regarded as a primary application for near-term quantum processors, although a definitive demonstration of a practical quantum advantage remains elusive. Recent studies have reported that both gate-based quantum circuits and quantum annealers can outperform state-of-the-art classical heuristics on multi-objective optimization (MO-MaxCut) problems. However, these studies did not fully account for the substantial pre- and postprocessing overheads intrinsic to quantum solvers, leading to incomplete comparisons between quantum and classical approaches. In this work, we re-examine the same benchmark suite using GPU-based quantum-annealing-inspired algorithms (QAIAs), which generate probabilistic samples analogously to quantum processors and thus serve as formidable classical contenders. 
%Our results show that QAIAs can sample candidate solutions approximately two orders of magnitude faster than the \REVdel{previously studied quantum processors} \REVadd{sampling and device-access times reported for the quantum processors studied previously}. 
Our results show that QAIAs can sample candidate solutions approximately two orders of magnitude faster than the previously studied quantum processors{, based on the reported sampling and device-access times}.
{On the instances for which a complete pipeline can be compared, the measured end-to-end QAIA runtime, including model construction, sampling and Pareto filtering, is about an order of magnitude shorter than that of the exact objective-space baselines DCM and DPA-a, and the QAIA pipeline outperforms the tested evolutionary multi-objective heuristics under an equal wall-clock budget.}

\end{abstract} %%%%%%%%%

% \bigskip

% \linenumbers
\twocolumn

\section*{Introduction}
Combinatorial optimization is a major application domain for near-term quantum and quantum-inspired computing, and a central testbed for assessing whether such platforms can offer practical advantages on hard discrete problems~\cite{Abbas2024}. This prominence is largely due to the fact that many industrial challenges—from logistics to financial modeling—can be mapped onto Ising or Quadratic Unconstrained Binary Optimization (QUBO) frameworks~\cite{Lucas2014, Glover2022}, which are natively compatible with quantum architectures. In the current era of Noisy Intermediate-Scale Quantum (NISQ) technology \cite{Preskill2018, bharti2022, Cheng2023}, hybrid algorithms and specialized hardware aim to find high-quality solutions to NP-hard problems more efficiently than classical counterparts.

Within this domain, multi-objective optimization (MOO)~\cite{Miettinen1998, ehrgott2005} addresses problems where multiple conflicting objectives must be optimized simultaneously. The goal of MOO is to identify Pareto-optimal solutions: trade-offs for which no objective can be improved without degrading at least one other. Their objective values form the Pareto front, the set of non-dominated points in objective space. Unlike single-objective optimization, MOO aims to approximate this front rather than return a single optimum. In this context, sampling-based solvers—such as gate-based quantum processors~\cite{Harrigan2021, Ebadi2022, Zhu2026}, quantum annealers \cite{Rajak2022, Kim2025, Albash2018}, quantum-inspired optimizers~\cite{Huang2023, Zeng2024}, and Ising machines~\cite{Mohseni2022}—are naturally relevant for the task. Indeed, unlike traditional iterative methods that isolate solutions sequentially, these samplers generate an ensemble of high-quality solutions, effectively mapping the Pareto front without the need for extensive computational cycles.

A preliminary study by Kotil et al.~\cite{Kotil2025} investigated the application of the Quantum Approximate Optimization Algorithm (QAOA)~\cite{Farhi2014, Blekos2024} to a specific class of MOO problems using IBM’s gate-based quantum hardware. Specifically, the authors demonstrated that gate-based systems could sample Pareto-optimal candidates for the three- and four-objective MO-MaxCut problem with 42 variables and hardware-native topology at speeds competitive with established classical algorithms.

Building upon this work, subsequent research by King~\cite{king2025multi} utilized Quantum Annealing (QA)~\cite{Kadowaki1998, Johnson2011} via D-Wave’s Advantage series hardware to address the same problem instances. This follow-up study reported that QA not only achieves higher sampling velocities but also requires a significantly lower sample volume to recover the complete Pareto set. As a result, QA can achieve the optimal hypervolume two to three orders of magnitude faster than the gate-based counterpart, highlighting its efficiency for these tasks.

However, these studies focused primarily on the time required to sample candidate solutions without accounting for the total execution time of the entire Pareto front generation process. A rigorous end-to-end comparison must encompass the model construction phase as well as the computationally intensive filtering of non-dominated solutions from the sampled set. Consequently, omitting these stages renders comparisons with classical baselines incomplete and may lead to an overestimation of the practical advantages offered by the quantum approaches.

In this work, we investigate the same problem instances using GPU-accelerated quantum-annealing-inspired algorithms ({QAIAs}), focusing on Simulated Bifurcation (SB)~\cite{goto2019, goto2021}. This framework enables us to evaluate modern classical Ising solvers on these benchmarks, with particular attention to the metrics critical for practical multi-objective optimization: sampling speed, sample diversity, and the efficiency of end-to-end Pareto-front recovery. 

QAIAs have previously demonstrated superior performance in solving single-objective MaxCut problems across diverse graph topologies~\cite{Huang2023, Zeng2024}. This motivates evaluating whether these algorithms exhibit similarly robust performance when extended to the multi-objective regime.

We show that GPU-accelerated QAIAs provide a strong classical baseline for multi-objective quantum-optimization studies: {in sampling time, they reach the hypervolume of the Pareto set reported in previous work roughly one order of magnitude faster than QA and four orders of magnitude faster than QAOA, where the comparison uses the sampling and device-access times reported in those works}. We further identify sample quality as a key factor in end-to-end efficiency, showing that QAIA recovers the {reported} Pareto set with about $10\times$ fewer candidates than QAOA, though with a few times more samples than QA on average. To improve Pareto-set exploration, we introduce a Gaussian-noise variant of SB that increases candidate diversity. Finally, evaluating the full pipeline---model construction, sampling, and Pareto filtering---we show that the QAIA framework {recovers the Pareto front reported in~\cite{Kotil2025}} about one order of magnitude faster than the {exact objective-space baselines DCM and DPA-a~\cite{BOLAND2017, Dachert2024} on the same instances}, while outperforming MOEA/D~\cite{zhang2007moea}, NSGA-II~\cite{deb2002fast}, NSGA-III~\cite{deb2013evolutionary,jain2013evolutionary,blank2019investigating}, and RVEA~\cite{cheng2016reference} in hypervolume and number of {recovered nondominated} solutions under {a} fixed wall-clock time budget on dense three-objective instances up to $n=200$.

These results indicate that, for the MO-MaxCut benchmarks considered here, GPU-accelerated quantum-inspired Ising solvers {provide a faster complete pipeline than the classical baselines tested here, and faster candidate sampling than the sampling and access times reported for the available quantum hardware}. They also demonstrate the potential of approximate Ising solvers for {complex MOO problems} and provide a quantitative reference point for future advances in quantum optimization.

{Our main contribution consists of (i) the observation that, in the multi-objective setting, sample diversity rather than single-objective solution quality is the binding constraint on end-to-end cost, and (ii) a pipeline-level benchmark with consistent timing conventions, which shows that Pareto filtering and sample diversity, rather than raw sampling speed, determine the practical cost of the whole workflow.}

All code related to our work is publicly available on GitHub at \url{https://github.com/Tao-qubit/QAIA-for-MOO}.

\section*{Hamiltonian formulation of multi-objective MaxCut}

Following Refs.~\cite{Kotil2025, king2025multi}, we address the multi-objective weighted maximum-cut problem, which allows a direct comparison between {QAIA-based solvers and} the methods reported in those studies on the same benchmark family. 
Consider an undirected graph $G=(V,E)$, where $V$ is a finite vertex set,
$E\subseteq \{(i,j): i,j\in V,\ i<j\}$ is the edge set, and $n=|V|$.

For each of the $K$ competing objectives, let $J_{ij}^{(k)}$ denote the weight of edge $(i,j)$ in objective $k$. Using spin variables $s_i \in \{-1,1\}$, the corresponding cut value can be written as
\begin{equation}
    \label{eq:single_obj}
    C_k(\mathbf{s}) = \frac{1}{2}\sum_{(i,j) \in E} J_{ij}^{(k)} \bigl(1 - s_i s_j\bigr).
\end{equation}
Here, an edge contributes when its endpoints are assigned opposite signs. Since the first term is constant, maximizing $C_k(\mathbf{s})$ is equivalent to minimizing the Ising Hamiltonian 
\begin{equation}
    H_k(\mathbf{s}) = \sum_{(i,j) \in E} J_{ij}^{(k)} s_i s_j.
\end{equation}
We therefore work directly with the Hamiltonian formulation, which is standard for Ising-based solvers \cite{Lucas2014}. {The constant term $\tfrac12\sum_{(i,j)\in E}J^{(k)}_{ij}$ is restored whenever cut values are reported, so all objective vectors and hypervolumes below are expressed in cut-value space.}

The MO-MaxCut problem is formulated on the configuration space $\mathcal S=\{-1,1\}^n$. Each configuration
$\mathbf{s}\in\mathcal S$ is assigned a vector of $K$ Ising Hamiltonians,
\begin{equation}
    \mathbf H(\mathbf s)
    =
    \bigl(H_1(\mathbf s),H_2(\mathbf s),\ldots,H_K(\mathbf s)\bigr)^{T}
    \in \mathbb R^K .
\end{equation}
Since the objectives are minimized simultaneously, solutions are compared
using the Pareto dominance order. A configuration $\mathbf{s}_1$ is said to
dominate $\mathbf{s}_2$, denoted $\mathbf{s}_1 \prec \mathbf{s}_2$, if
\begin{equation}
\begin{aligned}
    H_i(\mathbf{s}_1) &\le H_i(\mathbf{s}_2)
    \quad \forall i\in\{1,\ldots,K\},\\
    \text{and } 
    H_j(\mathbf{s}_1) &< H_j(\mathbf{s}_2)
    \quad \text{for some } j\in\{1,\ldots,K\}.
\end{aligned}
\end{equation}
Equivalently, $\mathbf H(\mathbf{s}_1)$ Pareto-dominates $\mathbf H(\mathbf{s}_2)$ in objective space. The goal is therefore not to minimize a scalar objective, but to identify the set of Pareto-optimal configurations
\begin{equation}
\mathcal P^\star
=
\bigl\{
\mathbf{s}\in\mathcal S : \ \
 \nexists\, \mathbf{s}'\in\mathcal S
\text{ such that } \\
\mathbf{s}' \prec \mathbf{s}
\bigr\}.
\end{equation}
The corresponding Pareto front $\mathcal F^\star$ is the image of this set in objective space. In practice, the sampler produces a finite set $\widehat{\mathcal P}\subseteq\mathcal S$ of nondominated spin configurations whose objective vectors approximate the Pareto front.

For evaluation, we report solution quality in the original cut-value space. Let $\mathbf C(\mathbf s) = \bigl(C_1(\mathbf s),\ldots,C_K(\mathbf s)\bigr)^T$.
The quality of the obtained approximation $\widehat{\mathcal P}$ is quantified using the \textit{hypervolume} (HV) indicator \cite{Zitzler2003, Afsar2023}, defined as
\begin{equation}
    HV(\widehat{\mathcal P}, \mathbf r)
    =
    \Lambda\left(
    \bigcup_{\mathbf s\in\widehat{\mathcal P}}
    [\mathbf r,\mathbf C(\mathbf s)]
    \right),
\end{equation}
where $\Lambda$ denotes the $K$-dimensional Lebesgue measure and $\mathbf r\in\mathbb R^K$ is a reference point dominated by all objective vectors in the approximation set, i.e., $r_k\le C_k(\mathbf s)$ for all $k$ and all $\mathbf s\in\widehat{\mathcal P}$. $[\mathbf{r}, \mathbf C(\mathbf{s})]=\prod_{k=1}^{K}[r_k,C_k(\mathbf s)]$ denotes the hyperrectangle spanned by the cut values of $\mathbf{s}$ and $\mathbf{r}$
(see Appendix~\ref{app:Hypervolume}).

\section*{Scalarization via the weighted sum method}

To solve the multi-objective problem using standard optimization solvers designed for single objectives, we employ the weighted sum method \cite{ehrgott2005}. For a weight vector $\mathbf{c} = [c_1, c_2, \dots, c_K]^T$, where $c_k \ge 0$ and $\sum_{k=1}^K c_k = 1$, the scalarized Hamiltonian is given by
\begin{equation}
    H_{\text{total}}(\mathbf{s}; \mathbf{c}) = \sum_{k=1}^{K} c_k H_k(\mathbf{s}) = \sum_{(i,j) \in E} J_{ij}(\mathbf{c}) s_i s_j,
    \label{eq:H_total}
\end{equation}
where $J_{ij}(\mathbf{c}) = \sum_{k=1}^{K} c_k J_{ij}^{(k)}$ is the effective coupling for edge $(i, j)$. Consequently, the original multi-objective problem is transformed into a family of single-objective minimization problems:
\begin{equation}
    \min_{\mathbf{s} \in \{-1, +1\}^n} H_{\text{total}}(\mathbf{s}; \mathbf{c}).
\end{equation}
By solving these problems for different vectors $\mathbf{c}$, we obtain candidate solutions that approximate the Pareto frontier of the original multi-objective problem. 

In the current work, we use a simplex-lattice design (Das–Dennis) \cite{Das_Dennis1998} to generate weight vectors. Specifically, for \(K\) objectives, we generate a uniform lattice (resolution \(H\)) on the \((K-1)\)-dimensional unit simplex \(\Delta^{K-1}\), i.e., all nonnegative weight vectors summing to one with components restricted to \(\{0,1/H,\dots,1\}\). This choice is well known in the area of multi-objective optimization \cite{Blank2020}; it produces evenly spaced vectors and allows better reproducibility of the results. 

Furthermore, we observed that the number of Pareto-optimal solutions obtained by an exact solver grows much slower than the number of vectors and seems to saturate well below 100 for the three-objective problem, while the full Pareto-set size is reported to be 2067 \cite{king2025multi}. This indicates that the majority of the Pareto-optimal solutions are not global optima for any weight vector {in the tested lattice}. We found that only a moderate number of well-spaced weight vectors is sufficient to allow the quantum-inspired solver to sample the Pareto set. {In our experiments we also remove extreme weight vectors lying on the vertices and edges of the simplex and only use the subset corresponding to interior points.} See more details in Appendix \ref{app:das-dennis}.

{Two caveats follow from the choice of scalarization. First, the weighted-sum method combined with an exact solver cannot reach Pareto-optimal points that are not optimal for any weight vector, i.e. the unsupported points in non-convex regions of the front~\cite{ehrgott2005}. The role of NI-SB is to add, besides near-optimal solutions of each subproblem, a diverse set of low-energy configurations that may populate such regions (Appendix~\ref{app:das-dennis}). Second, the number of lattice points grows combinatorially, $\binom{H+K-1}{K-1}$, so the number of subproblems, the memory requirement, and the filtering cost all increase rapidly with $K$; this is one reason the advantage is smaller for four objectives than for three.}

\begin{table*}[]
\resizebox{\linewidth}{!}{
\begin{tabular}{@{}cccc@{}}
\toprule
Method            & Type      & Hardware information and details & Source                                               \\ \midrule
QA (Advantage) & Quantum   & D-Wave Advantage System 4.1, measured access time  & \cite{king2025multi}      \\
QA (Advantage2)  & Quantum   & D-Wave Advantage2 System 1.6, measured access time  & \cite{king2025multi}   \\
QAOA (IBM\_Fez) v1   & Quantum   & 156-qubit IBM\_Fez, $p=6$, assuming 10,000 shots/s  &  \cite{Kotil2025}                            \\
QAOA (IBM\_Fez) v2   & Quantum   & 156-qubit IBM\_Fez, $p=6$, fidelity estimate 3.71\% assuming 10,000 shots/s &  \cite{Kotil2025} \\
QAOA (IBM\_Fez) v3   & Quantum   & 156-qubit IBM\_Fez, $p=6$, fidelity estimate 53.0\% assuming 10,000 shots/s & \cite{Kotil2025} \\
QAOA (MPS-$\chi=50$)   & Classical & MPS simulation on Apple M1 Max CPU, bond dimension $\chi=50$  &   \cite{Kotil2025}  \\
QAOA (MPS-$\chi=20$)   & Classical & MPS simulation on Apple M1 Max CPU, bond dimension $\chi=20$   &  \cite{Kotil2025}        \\
DCM and DPA-a     & Classical & AMD EPYC 7773X CPU      &  \cite{Kotil2025}                                                     \\
NI-bSB and NI-dSB & Classical & AMD EPYC 7773X CPU and NVIDIA GeForce RTX 4090 GPU   & This work        \\ \bottomrule
\end{tabular}
}
\caption{Hardware information for methods shown in Fig.~\ref{fig:3&4obj}. {The QA entries are measured QPU access times; the QAOA v1 entry is based on an assumed shot rate, and the v2 and v3 entries are fidelity-rescaled projections built on that assumption. The DCM and DPA-a entries are exact objective-space algorithms.}}
\label{tab:hardware}
\end{table*}

\begin{figure*}[t]
    \centering
    \includegraphics[width=1\textwidth]{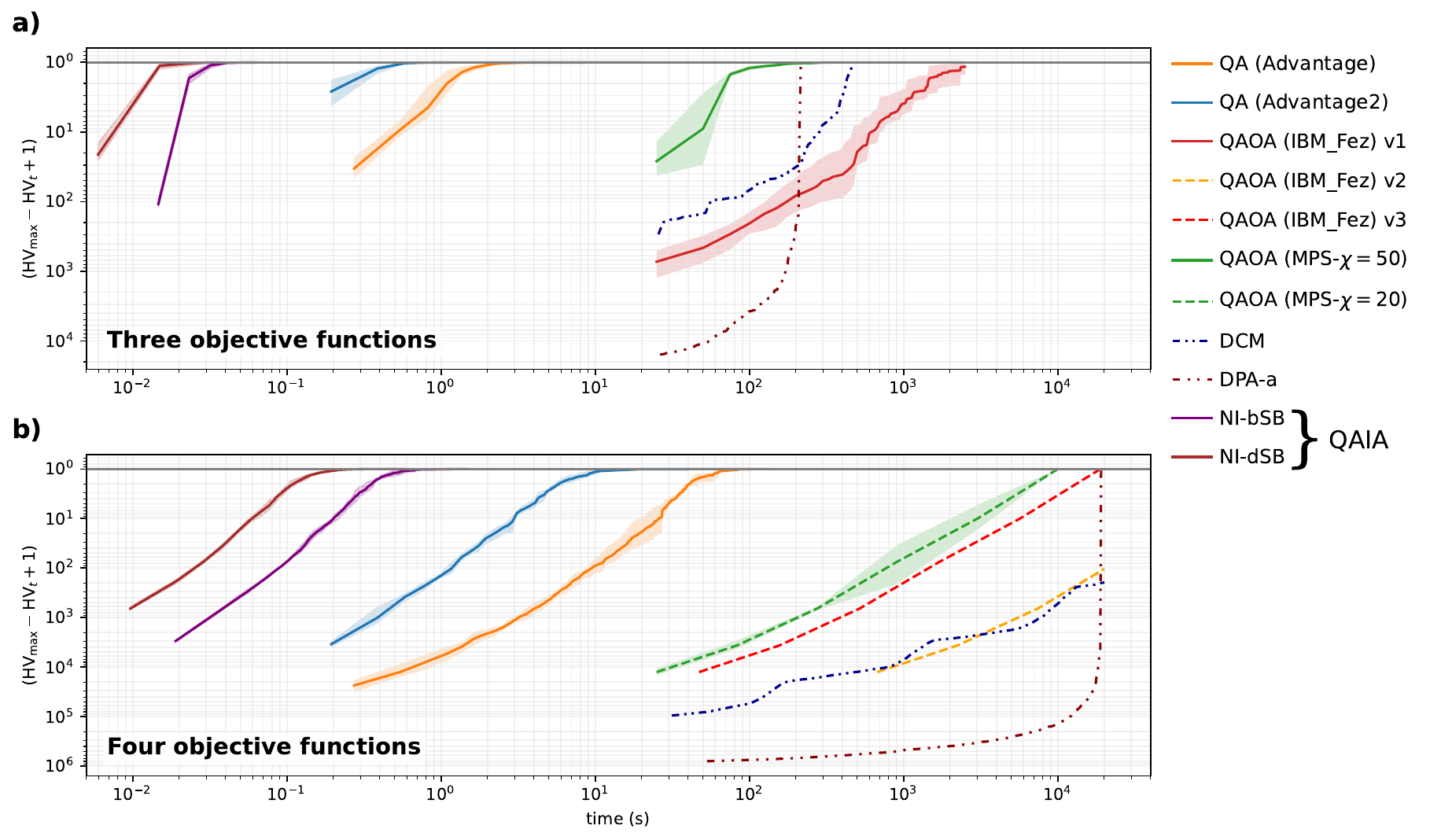}
    \caption{Convergence of different algorithms toward the Pareto front. The y-axis shows the Hypervolume difference ($HV_{\text{max}} - HV_t + 1$) on a logarithmic scale, and the x-axis represents the sampling time in seconds. Data for all algorithms except SB are taken from Refs.~\cite{Kotil2025, king2025multi}. See the details and hardware information in Table \ref{tab:hardware}. {Curves for NI-bSB and NI-dSB are means over five independent runs, and the shaded regions span the minimum--maximum range over those runs. For QA and QAOA only sampling or device-access times are available in the source publications, so this panel does not compare end-to-end runtimes.}}
    \label{fig:3&4obj}
\end{figure*}

\section*{Pareto filtering}

For an Ising-based multi-objective solver, the sampler does not return a
Pareto set directly. For a scalarization vector $\boldsymbol c$, the
sampled Hamiltonian has the form introduced in Eq.~(\ref{eq:H_total}),
and repeated runs produce a finite pool of candidate spin configurations.
The union of samples obtained over one or several scalarizations must
therefore be converted into a nondominated approximation set by a classical
postprocessing step.

Let
\[
\mathcal X=\{\mathbf s^{(1)},\ldots,\mathbf s^{(M)}\}
\]
denote the raw sample pool. Pareto filtering evaluates the objective vector
$\mathbf C(\mathbf s)$ for each $\mathbf s\in\mathcal X$, removes duplicate
configurations or duplicate objective vectors if required, and retains only
those configurations that are not dominated by any other sampled
configuration. In cut-value maximization space, this yields
\[
\begin{aligned}
   \widehat{\mathcal P}
    =
    \big\{
    \mathbf s\in\mathcal X:
    \nexists\,\mathbf s'\in\mathcal X
    \text{ such that }
    \mathbf{s}' \prec \mathbf{s}
    % C_k(\mathbf s')\ge C_k(\mathbf s)\ \forall k \\
    % \text{ and }
    % C_j(\mathbf s')>C_j(\mathbf s)\ \text{for some }j
    \big\}. 
\end{aligned}
\]
The corresponding approximation front is then
$\widehat{\mathcal F}=\{\mathbf C(\mathbf s):\mathbf s\in\widehat{\mathcal P}\}$.

This filtering step is deterministic but can be computationally nontrivial, contributing significantly to the end-to-end runtime.
A naive nondominated sorting procedure requires pairwise comparison of sampled configurations and has time complexity $O(KM^2)$ for $M$ samples and $K$ objectives. If many repeated configurations or repeated objective vectors are produced, deduplication reduces the effective input size; otherwise, filtering nondominated solutions becomes a computational bottleneck. For this reason, we treat Pareto filtering as a separate postprocessing stage in the computational pipeline and evaluate its overhead explicitly. {In our implementation the filter is provided by the moocore library\cite{moocore}, and deduplication together with the symmetry reduction above reduces the effective input size before the filter is applied. We report the measured filtering cost rather than a fitted scaling law; the $O(KM^2)$ bound refers to the naive pairwise procedure and should be read as a worst-case estimate.}

\section*{Simulated Bifurcation with Gaussian noise}

To solve the composite Ising models constructed in the previous section, we employ the {Simulated Bifurcation (SB)} algorithm \cite{goto2019, goto2021}. {SB is chosen as the representative QAIA because it admits an efficient batched GPU implementation and has been reported to perform well on single-objective MaxCut across graph topologies~\cite{Huang2023,Zeng2024}; comparisons with SimCIM and simulated annealing, are reported in Appendix~\ref{app:simcim}.} For each weight vector $\mathbf{c}$, we run SB on the scalarized Ising matrix $J_{ij}(\mathbf{c})$; in the equations below, $J_{ij}$ denotes this scalarized coupling matrix. SB is a heuristic algorithm inspired by the adiabatic evolution of quantum systems but implemented entirely on classical hardware by simulating classical Hamiltonian dynamics. It simulates a network of nonlinear oscillators that undergo bifurcations to converge toward low-energy states of the Ising Hamiltonian. In this work, we use two primary variants: {ballistic SB (bSB)} and {discrete SB (dSB)}, and introduce a stochastic noise term to enhance sampling diversity for multi-objective optimization. A similar noise term is present in the SimCIM algorithm \cite{tiunov2019a}. In SimCIM it is used to escape local minima and improve single-objective solution quality; here, we use it to increase sample diversity for better coverage of the Pareto frontier. Overall, we found that SimCIM is less effective in this setting, see Appendix~\ref{app:simcim}.

In SB, each Ising spin $s_i \in \{-1,+1\}$ is represented by a continuous “soft spin” variable $x_i \in [-1,1]$ and its conjugate momentum $y_i$. The system evolves according to the following ordinary differential equations:
\begin{equation}
\begin{aligned}
    \dot{x}_i &= a_0 y_i, \\
    \dot{y}_i &= -[a_0 - a(t)]x_i + c_0 \left( \sum_{j=1}^n J_{ij} \phi(x_j) \right),
\end{aligned}
\end{equation}
where $a(t)$ increases linearly from 0 to $a_0=1$ over time, driving a bifurcation: initially all trajectories converge to the origin, but as $a(t)$ exceeds a critical threshold, the origin becomes unstable and $x_i$ diverge toward $\pm 1$. Throughout this work, $a_0$ is fixed to $1$ and the coupling scale is normalized as $c_0 = 1 / \max_i |\sum_j J_{ij}|$. The function $\phi(x_j) = x_j$ defines bSB (ballistic SB), while $\phi(x_j) = \mathrm{sgn}(x_j)$ defines dSB (discrete SB). The final spin assignment is $s_i = \mathrm{sgn}(x_i)$.

To overcome the limited diversity of deterministic SB—which often converges to the same solution across runs for a fixed objective weight vector—we introduce Noise-Injected Simulated Bifurcation (NI-SB). We augment the momentum equation with {a Gaussian random term}:
\begin{equation}
    \label{eq:NI-SB_equation}
    \dot{y}_i = -[a_0 - a(t)]x_i + c_0 \left( \sum_{j=1}^n J_{ij} \phi(x_j) \right) + \alpha \eta_i,
\end{equation}
where $\eta_i \sim \mathcal{N}(0,1)$ is independently sampled for each spin at every integration step, and $\alpha > 0$ controls noise intensity. This perturbation acts as synthetic thermal noise, enabling trajectories to explore a diverse set of solution candidates. In the context of MOO, this mechanism significantly increases the probability of discovering distinct Pareto-optimal solutions for the same weight vectors $\{\mathbf{c}^{(i)}\}$, thereby enriching the Pareto front approximation, as illustrated in Fig.~\ref{fig:sample_advantage}(a).

Starting from {random initial values drawn uniformly from $[-0.005,0.005]$}, the dynamical equations are integrated using an explicit Euler scheme with time step $\text{d}t=1$. To enforce $x_i \in [-1,1]$, an inelastic wall boundary condition is applied: whenever $|x_i| > 1$, we project $x_i \leftarrow \mathrm{sgn}(x_i)$ and reset $y_i \leftarrow 0$, mimicking a perfectly inelastic collision.

{\textbf{Timing conventions.} We use two timing quantities throughout. \emph{Sample-generation time} is the wall-clock time spent generating candidates, i.e. inside the SB update, excluding model construction and postprocessing; it is the counterpart of the sampling and QPU-access times reported for QA and QAOA and is the quantity shown in Fig.~\ref{fig:3&4obj}. \emph{End-to-end pipeline time} is the wall-clock time of the complete solve on our host: construction of the scalarized coupling matrix, sampling, and Pareto filtering; it is the quantity shown in Fig.~\ref{fig:3&4obj_ETE} and used in Table~\ref{tab:MOEA_compare}. Each reported point is the cumulative wall-clock time within one run, measured with Python's \texttt{time.time()}. }

\section*{Results and discussion}

To evaluate the performance of the proposed Noise-Injected Simulated Bifurcation (NI-SB) framework, we conducted experiments on MO-MaxCut problems and compared the results against state-of-the-art quantum and classical solvers. The performance is assessed based on two key metrics: the speed of sampling candidate solutions for Pareto front approximation, and end-to-end performance of approximating the Pareto front measured by relative hypervolume and solution time.

\subsection*{Sampling time advantage}

Fig.~\ref{fig:3&4obj} illustrates the sampling performance of various algorithms. A full list of compared methods is presented in Table \ref{tab:hardware}. The comparison includes QA executed on two generations of D-Wave Advantage systems, QAOA executed on IBM Fez QPU, MPS simulations of QAOA on classical hardware, classical exact algorithms (DCM, DPA-a), and our proposed NI-bSB/NI-dSB. 

We generated 190 evenly distributed Das–Dennis weight vectors for the three-objective problem and 220 for the four-objective problem. Each SB call was run with 50 iterations, a batch size of {3,000 samples per weight vector,} and a noise amplitude of 0.15 or 0.1 for the three- and four-objective problem, respectively. SB algorithms were called repeatedly until covering all Pareto-optimal solutions reported in previous works. On a GPU with $16$ GB of memory, for a sparse problem with only 42 variables it is possible to {generate the full 3,000-sample batch for every weight vector in a single parallel call}, using half-precision (float16) arithmetic.

\begin{figure*}[t]
    \centering
    \includegraphics[width=1\textwidth]{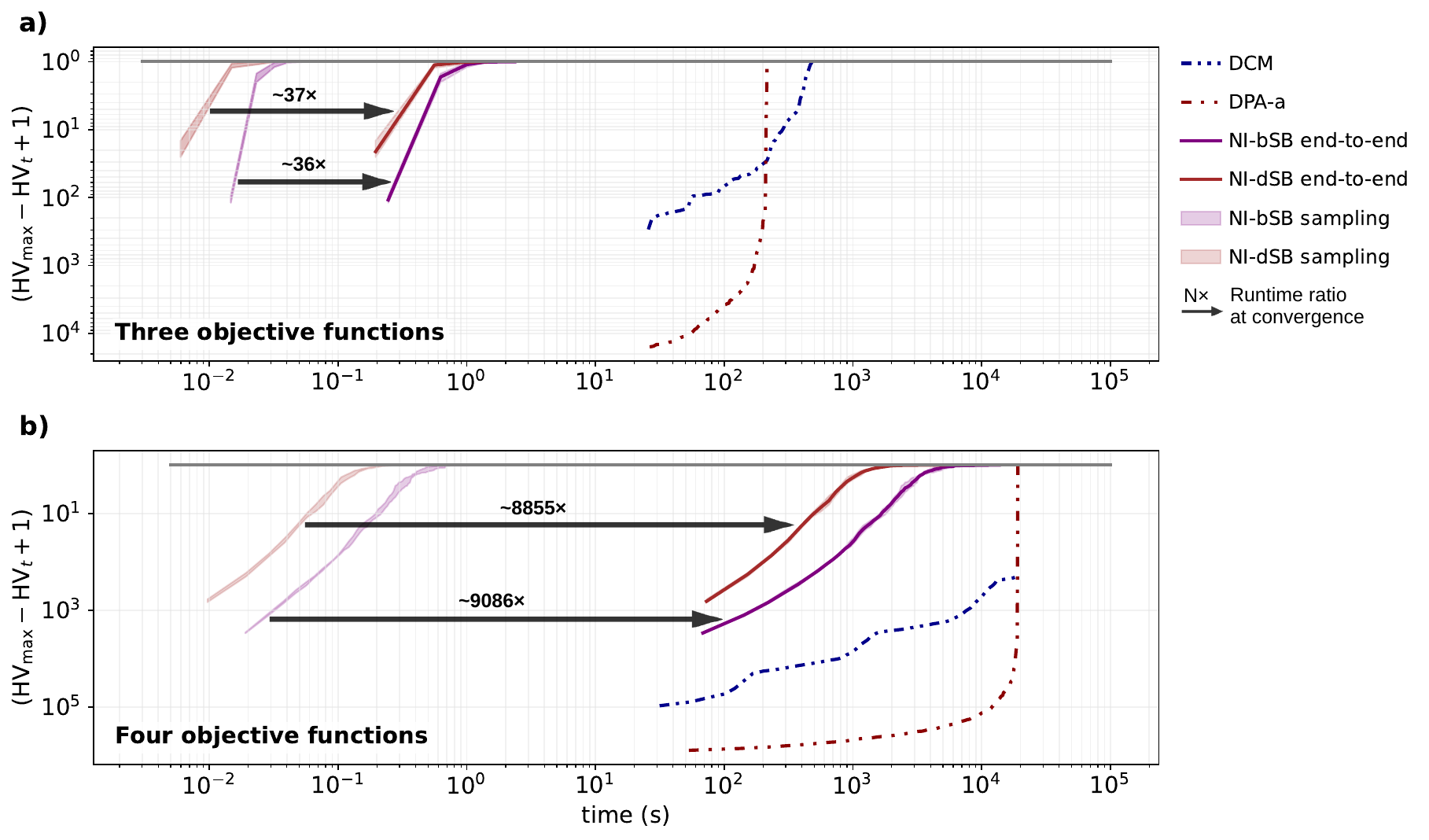}
    \caption{Comparison of sampling time with full Pareto front reconstruction time. The y-axis shows the Hypervolume difference ($HV_{\text{max}} - HV_t + 1$) in logarithmic scale, and the x-axis represents the end-to-end time in seconds. DCM and DPA-a are {exact objective-space} classical baseline algorithms. End-to-end time for NI-dSB and NI-bSB includes both model construction and filtering of Pareto-optimal solutions from the sampled candidate set. {The annotations give the ratio of end-to-end to sampling-only time for each variant; the sampling-only curves are reproduced from Fig.~\ref{fig:3&4obj} for reference. Curves are means over five independent runs and the shaded regions span the minimum--maximum range over those runs. }
    }
    \label{fig:3&4obj_ETE}
\end{figure*}

As shown in Fig.~\ref{fig:3&4obj}, the NI-SB algorithms (purple and brown lines) exhibit a substantial advantage in speed of sampling the candidate set that contains the Pareto-optimal solutions. Both bSB and dSB reach the {best-known} hypervolume plateau within approximately $10^{-2}$ to $10^{-1}$ seconds. In contrast, QA systems require several seconds to minutes to achieve comparable performance, while QAOA and baseline classical methods are orders of magnitude slower. This result shows that leveraging GPU-accelerated classical dynamics allows for extremely rapid sampling of high-quality solutions, making NI-SB highly suitable for time-critical applications.

\subsection*{Sample quality and end-to-end solution time}

Although pure sampling-time efficiency is promising, it is the end-to-end time that defines practical algorithmic utility. This time includes not only the candidate-sampling time reported in \cite{Kotil2025, king2025multi} and shown in Fig.~\ref{fig:3&4obj}, but also other sources of delay, including circuit compilation and parameter training for QAOA and—importantly—sample postprocessing to define nondominated solutions that constitute the Pareto front. Although all heuristic solvers deliver samples that contain Pareto-optimal solutions, postprocessing is still needed to define those solutions within the sample. This postprocessing time {grows with the number of distinct candidates} and may take a few seconds to a few minutes for millions of samples, as found for all three algorithms. This makes sample quality, i.e. the ratio of total sample size to the Pareto-set size, a crucial indicator of end-to-end performance. 

\begin{figure*}[ht]
    \centering
    \includegraphics[width=1\textwidth]{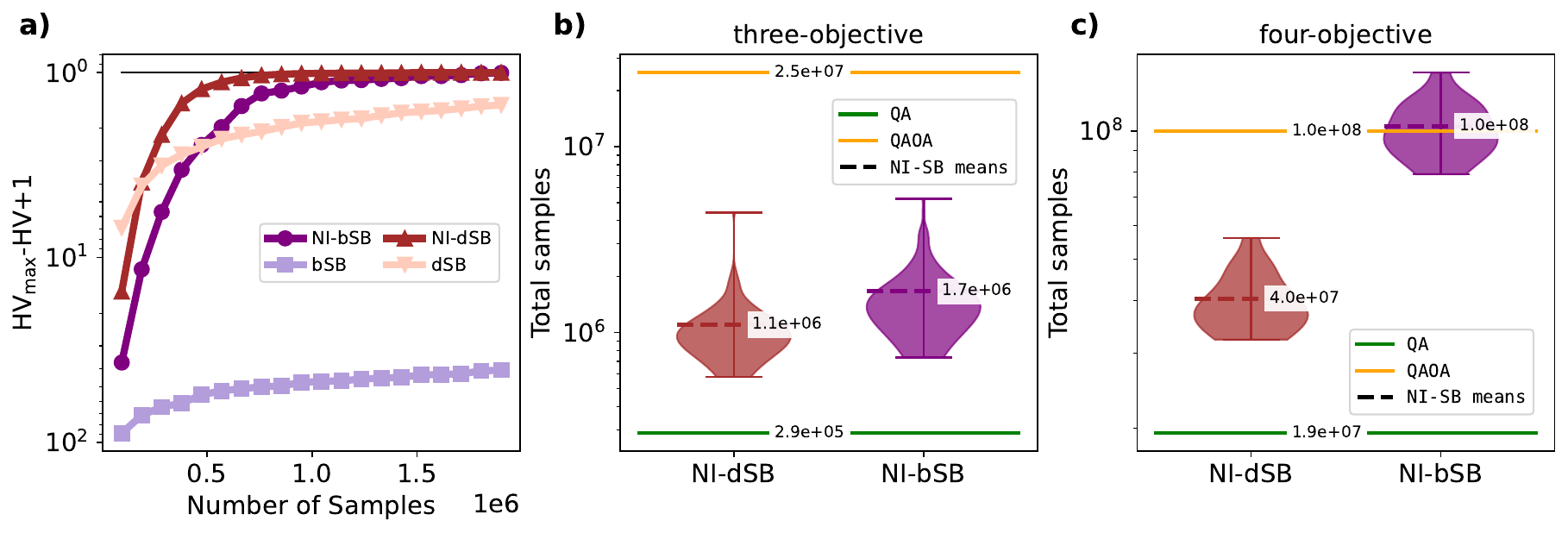}
    \caption{\textbf{a)} Convergence to {the best-known} hypervolume for original and noisy SB. \textbf{b)} and \textbf{c)} Distribution of numbers of samples needed by NI-SB to discover the {reported} Pareto set of the three- and four-objective problems, respectively. Values for QA and QAOA are taken from \cite{king2025multi} and \cite{Kotil2025}, respectively. Note that QAOA result for four-objective problem is approximate, as it includes 10 nondominated solutions less than results of QA and NI-SB. Thus the plotted QAOA value in four-objective case should be interpreted as a lower bound on the number of samples required for complete front recovery. {Panel (a) shows means over five independent runs; panels (b) and (c) show distributions over 100 independent trials, with the horizontal lines marking the means. The target is the hypervolume of the Pareto set reported in Refs.~\cite{king2025multi,Kotil2025}.}}
    \label{fig:sample_advantage}
\end{figure*}

In Fig.~\ref{fig:sample_advantage}~(b)~and~(c) we show that the total number of samples required by NI-SB to find the {reported} Pareto set lies between the numbers reported in previous works for QAOA and QA. This suggests that sampling solutions with NI-SB should reduce postprocessing time compared to QAOA, while possibly remaining less sample-efficient than QA.

In Fig.~\ref{fig:3&4obj_ETE} we show the end-to-end time taken by our NI-SB-based solver, including Ising model construction, sampling, and sample postprocessing to extract the Pareto set. We employ Fast Non-dominated Sorting for Pareto front filtering in the postprocessing stage, as implemented in the moocore Python package~\cite{moocore}. We compare this time to the execution time of classical baseline algorithms run on the same hardware. The results show that the QAIA-based solver is about an order of magnitude faster than those classical algorithms. Our analysis suggests that QA on moderate-size hardware-native graphs may show even better end-to-end performance{, owing to its higher sample efficiency (Fig.~\ref{fig:sample_advantage})}.

\subsection*{Comparison to evolutionary algorithm-based heuristics}
In the previous sections, we benchmarked our approach against various Ising solvers in terms of sampling speed and compared it end-to-end with classical exact algorithms. In this section, we further evaluate its performance against MOEA/D~\cite{zhang2007moea}, NSGA-II~\cite{deb2002fast}, NSGA-III~\cite{deb2013evolutionary,jain2013evolutionary,blank2019investigating}, and RVEA~\cite{cheng2016reference}, as implemented in Pymoo \cite{pymoo}. These algorithms are well-established and widely adopted benchmarks in the multi-objective optimization field, representing distinct and influential methodological paradigms.

Drawing on the methodology in~\cite{DeSantis2026}, we randomly generated three-objective MaxCut instances such that the cut value $C_1$ exhibits a strong negative correlation with $C_2$, with Pearson's coefficient $\approx -0.7$. The third objective $C_3$ is nearly uncorrelated with both $C_1$ and $C_2$, with Pearson's coefficient $\approx 0$. See more details on problem instance generation in Appendix \ref{app:3D_front}.

All algorithms are evaluated with a runtime limit of 5 seconds for instances with $n=25,42$, and 10 seconds otherwise. MOEA/D, NSGA-II, NSGA-III, and RVEA each use a population size of 190 and are terminated when reaching the runtime limit. The Das-Dennis method is used when reference directions are needed. Standard binary optimization settings are applied, including BinaryRandomSampling, TwoPointCrossover, and BitflipMutation, with GPU-accelerated function evaluations. Similarly, bSB and dSB utilize 190 Das-Dennis weight vectors and are configured with a noise amplitude of 0.1. For instances with $n=25,42$, the QAIA run for 50 iterations; otherwise, they run for 1000 iterations. Both algorithms generate batches of 570,000 candidate solutions in parallel for all weight vectors, followed by Pareto filtering. These sampling-filtering steps are repeated until the runtime limit is reached. A set of solutions filtered from randomly sampled candidate set is included as sanity baseline, subject to the same runtime limit.

We report both the number of recovered Pareto-front points and the hypervolume ratio achieved by each algorithm. 
Specifically, for instances with $n=25$, the optimal Pareto fronts were obtained by brute force and the reference points for hypervolume calculation were determined by the minimum cut values of each objective. For the IBM data~\cite{Kotil2025}, the reference point and reference front were adopted from~\cite{Kotil2025, king2025multi}.
For larger instances with $n=100$ and $n=200$, {a best-known composite archive was formed by pooling all nondominated solutions found by the compared algorithms, and the reference point was set from the extreme values observed in the pooled data.} We used 7 instances with varying sizes and densities. For each instance, every algorithm was executed for five independent runs for evaluation, and average values are presented in Table~\ref{tab:MOEA_compare}.

Brute-forcing an instance with a solution space of $2^{25}$ requires approximately 70 seconds. In contrast, within only 5 seconds, both QAIA and random sampling can generate roughly ten million samples and perform filtering, with the latter dominating the computational cost as shown in Fig.~\ref{fig:3&4obj_ETE}. Using this sample size, random sampling recovers over 80\% of the optimal hypervolume, while QAIA captures the entire Pareto front. Evolutionary algorithms, despite producing significantly fewer solutions, still achieve more than 90\% of the optimal hypervolume.

All experiments were conducted on a machine equipped with an AMD EPYC 7773X CPU and an NVIDIA GeForce RTX 4090 GPU. 

\begin{table*}[ht]
\resizebox{\linewidth}{!}{
\begin{tabular}{cccc|cc|cc|cc|cc|cc|cc|cc}
\hline
\multicolumn{4}{c|}{Instance} & \multicolumn{2}{c|}{Random} & \multicolumn{2}{c|}{RVEA} & \multicolumn{2}{c|}{MOEA/D} & \multicolumn{2}{c|}{NSGAII} & \multicolumn{2}{c|}{NSGAIII} & \multicolumn{2}{c|}{NI-bSB} & \multicolumn{2}{c}{NI-dSB} \\ \hline
$n$ & $d$ & \begin{tabular}[c]{@{}c@{}}Ref. front \\ size\end{tabular} & \begin{tabular}[c]{@{}c@{}}Reference\\ type\end{tabular} & \begin{tabular}[c]{@{}c@{}}Front \\ points\end{tabular} & \begin{tabular}[c]{@{}c@{}}HV\\ Ratio\end{tabular} & \begin{tabular}[c]{@{}c@{}}Front \\ points\end{tabular} & \begin{tabular}[c]{@{}c@{}}HV\\ Ratio\end{tabular} & \begin{tabular}[c]{@{}c@{}}Front \\ points\end{tabular} & \begin{tabular}[c]{@{}c@{}}HV\\ Ratio\end{tabular} & \begin{tabular}[c]{@{}c@{}}Front \\ points\end{tabular} & \begin{tabular}[c]{@{}c@{}}HV\\ Ratio\end{tabular} & \begin{tabular}[c]{@{}c@{}}Front \\ points\end{tabular} & \begin{tabular}[c]{@{}c@{}}HV\\ Ratio\end{tabular} & \begin{tabular}[c]{@{}c@{}}Front \\ points\end{tabular} & \begin{tabular}[c]{@{}c@{}}HV\\ Ratio\end{tabular} & \begin{tabular}[c]{@{}c@{}}Front \\ points\end{tabular} & \begin{tabular}[c]{@{}c@{}}HV\\ Ratio\end{tabular} \\ \hline
25 & 0.5 & 646 & \multirow{2}{*}{\begin{tabular}[c]{@{}c@{}}True\\ front\end{tabular}} & 14 & 86.5\% & 30 & 87.9\% & 41 & 68.1\% & 123 & 95.2\% & 108 & 95.3\% & \textbf{646} & \textbf{100\%} & \textbf{646} & \textbf{100\%} \\
25 & 1.0 & 649 &  & 17 & 86.5\% & 24 & 87.1\% & 35 & 67.2\% & 120 & 95.3\% & 80 & 95.0\% & \textbf{649} & \textbf{100\%} & \textbf{649} & \textbf{100\%} \\ \hline
42 & \begin{tabular}[c]{@{}c@{}}heavy-hex\\ \cite{Kotil2025}\end{tabular} & 2067 & \begin{tabular}[c]{@{}c@{}}Best-found \\ \cite{Kotil2025, king2025multi}\end{tabular} & 0 & 71.3\% & 12 & 91.9\% & 4 & 83.9\% & 34 & 95.3\% & 51 & 95.1\% & \textbf{2067} & \textbf{100\%} & \textbf{2067} & \textbf{100\%} \\ \hline
100 & 0.5 & 5694 & \multirow{4}{*}{\begin{tabular}[c]{@{}c@{}}Composite\\set\end{tabular}} & 0 & 27.2\% & 0 & 55.9\% & 0 & 36.8\% & 7 & 62.4\% & 24 & 76.4\% & 2998 & 97.9\% & \textbf{4016} & \textbf{99.4\%} \\
100 & 1.0 & 5123 &  & 0 & 25.4\% & 2 & 60.1\% & 1 & 38.4\% & 2 & 58.2\% & 17 & 74.8\% & 2702 & 97.9\% & \textbf{3541} & \textbf{99.4\%} \\
200 & 0.5 & 3740 &  & 0 & 17.9\% & 18 & 67.1\% & 0 & 28.4\% & 14 & 57.1\% & 48 & 75.0\% & 1519 & 93.1\% & \textbf{2451} & \textbf{98.4\%} \\
200 & 1.0 & 1996 &  & 0 & 19.2\% & 23 & 66.3\% & 3 & 30.5\% & 62 & 64.7\% & 69 & 83.5\% & 746 & 84.9\% & \textbf{1223} & \textbf{93.4\%} \\ \hline
\end{tabular}
}
\caption{Performance comparison of RVEA, MOEA/D, NSGA-II, NSGA-III, NI-bSB, and NI-dSB on three-objective MaxCut instances with varying node count ($n$) and edge density ($d$). Random uniform sampling is included to serve as the sanity baseline. For each algorithm, we report the mean {over five independent runs of the} number of recovered reference-front solutions and {of} the achieved hypervolume ratio. All experiments were conducted on a machine equipped with an AMD EPYC 7773X CPU and an NVIDIA GeForce RTX 4090 GPU.
}
\label{tab:MOEA_compare}
\end{table*}

The largest instance considered here is the complete graph on $n=200$ vertices. This exceeds the reported $180-193$ node estimate for the maximum clique size embeddable into Pegasus-16 and Zephyr-12 hardware graphs of the current D-Wave Advantage and Advantage2 quantum annealers \cite{Quinton2025, dwave_docs_advantage2_2025}. Even if these numbers improve as hardware progresses, {a} significant increase in the number of qubits {would be required to exploit} the advantage in parallel execution demonstrated in Ref.~\cite{king2025multi}.

\section*{Conclusions}

We presented an efficient multi-objective optimization framework based on Quantum-Annealing-Inspired Algorithms ({QAIAs}). The core of our approach uses a Gaussian-noise variant of Simulated Bifurcation (SB) to enhance sample diversity and mitigate premature convergence. We evaluated the method on MO-MaxCut instances previously used in quantum-computing studies { and on synthetic dense instances}.

Our results complement recent work on Ising-based approaches to multi-objective optimization~\cite{Kotil2025, king2025multi} by providing a strong (quantum-inspired) classical baseline. In numerical experiments, the proposed method generates high-quality samples substantially faster than QA as reported in \cite{king2025multi}, and recovers the {reported} Pareto set of three-objective MO-MaxCut with fewer candidate samples than {the QAOA estimate reported there}, although it still requires more samples than QA on average{. As noted in Fig.~\ref{fig:sample_advantage}(c), the four-objective QAOA sample count is a lower bound, because that reference front contains ten fewer nondominated solutions; the same caveat applies to the four-objective comparisons quoted here}.

In full end-to-end evaluations that include model construction, candidate sampling, and Pareto filtering, the method outperforms the considered classical heuristic baselines in hypervolume and number of recovered nondominated solutions under {the same wall-clock budget} on synthetic MO-MaxCut instances. These findings suggest that performance advantages previously observed for QAIA in some single-objective settings {can also appear} in multi-objective optimization.

On the benchmark instances from prior quantum studies, the full pipeline is faster than the {exact} DCM and DPA-a baselines reported in Ref. \cite{Kotil2025}. However, the advantage decreases as the number of objectives increases from three to four.

{Although no complete end-to-end quantum timing can be reconstructed from the published results, the reported quantum sampling and device-access times are slower than the QAIA sampling times on these benchmarks.} {The quantum results} remain relevant because they belong to the same class of approximate Ising solvers. The higher sample efficiency reported for QA in~\cite{king2025multi} is particularly interesting and suggests that sample quality should be studied independently from sampling speed.

Future work should develop quantitative indicators that separate sampling speed from sample quality, such as nondominated yield, Pareto-set recall, diversity, and hypervolume gain per sample. It will also be important to test the approach beyond MO-MaxCut on a broader range of multi-objective optimization problems{, including constrained problems, where penalty scaling and feasibility handling enter, and problems with integer or continuous variables that require a binary encoding}. {A supportedness analysis on instances with known fronts, together with a scaling study over $n$, graph density, and $K$, would further clarify when the approach is preferable.} Finally, other QAIA variants, such as thermally assisted SB~\cite{Kanao2022}, GSB~\cite{Xiao2026}, and Tabu-SB~\cite{Tao2026}, should be examined for possible additional benefits.

\section*{Acknowledgments}

We thank Marianna De Santis and Yue Zhang for their valuable discussions on the Multi-Objective Max-Cut problem. We also thank our colleague Zhuang Jiapei for sharing his valuable insights on implementation of classical MOO algorithms.

This work is supported by the Shenzhen International Quantum Academy (Grant No.SIQA2024KFKT01).

% \red{
% \section*{Data Availability}
% The datasets and source codes generated and analyzed during the current study are available in the GitHub repository:
% \url{https://github.com/Tao-qubit/QAIA-for-MOO}.
% The benchmark datasets used in this work include publicly available instances from:
% \url{https://github.com/andrew-d-king/QuantumAnnealingMOO}.
% These resources provide the necessary data and implementation details to reproduce the results reported in this paper.}

\section*{Data and code availability}

{The code, the benchmark instance files, and the raw timing data that support the findings of this study are available at \url{https://github.com/Tao-qubit/QAIA-for-MOO}.}

\section*{Use of generative AI}

During the preparation of this work, the authors used generative AI tools, including \mbox{Qwen}, \mbox{Kimi}, \mbox{ChatGPT}, and \mbox{NotebookLM}, to support literature search, text drafting and refinement, code prototyping, and code improvement. Parts of the manuscript were refined using the AI-enabled \LaTeX{} editor Prism. These tools were used only as assistance in the preparation process and not as substitutes for scientific judgment. All AI-assisted outputs, including text, code, references, data, analyses, and interpretations, were reviewed, verified, and revised as necessary by the authors. The authors take full responsibility for the accuracy, integrity, claims, results, conclusions, and overall content of this work.

% \clearpage
% \newpage

\bigskip
\bibliographystyle{vancouver}
\bibliography{references}

\begin{thebibliography}{10}

\bibitem{Abbas2024}
Abbas A, Ambainis A, Augustino B, B{\"a}rtschi A, Buhrman H, Coffrin C, et~al.
\newblock Challenges and opportunities in quantum optimization.
\newblock Nature Reviews Physics. 2024 Dec;6(12):718-35.
\newblock Available from: \url{https://doi.org/10.1038/s42254-024-00770-9}.

\bibitem{Lucas2014}
Lucas A.
\newblock Ising formulations of many NP problems.
\newblock Frontiers in Physics. 2014;Volume 2 - 2014.
\newblock Available from: \url{https://www.frontiersin.org/journals/physics/articles/10.3389/fphy.2014.00005}.

\bibitem{Glover2022}
Glover F, Kochenberger G, Du Y.
\newblock In: Punnen AP, editor. Applications and Computational Advances for Solving the QUBO Model. Cham: Springer International Publishing; 2022. p. 39-56.
\newblock Available from: \url{https://doi.org/10.1007/978-3-031-04520-2_2}.

\bibitem{Preskill2018}
Preskill J.
\newblock Quantum {C}omputing in the {NISQ} era and beyond.
\newblock {Quantum}. 2018 Aug;2:79.
\newblock Available from: \url{https://doi.org/10.22331/q-2018-08-06-79}.

\bibitem{bharti2022}
Bharti K, Cervera-Lierta A, Kyaw TH, Haug T, Alperin-Lea S, Anand A, et~al.
\newblock Noisy intermediate-scale quantum algorithms.
\newblock Reviews of Modern Physics. 2022;94(1):015004.

\bibitem{Cheng2023}
Cheng B, Deng XH, Gu X, He Y, Hu G, Huang P, et~al.
\newblock Noisy intermediate-scale quantum computers.
\newblock Frontiers of Physics. 2023 Mar;18(2):21308.
\newblock Available from: \url{https://doi.org/10.1007/s11467-022-1249-z}.

\bibitem{Miettinen1998}
Miettinen K.
\newblock Nonlinear Multiobjective Optimization.
\newblock 1st ed. International Series in Operations Research and Management Science. Springer New York; 1998.

\bibitem{ehrgott2005}
Ehrgott M.
\newblock Multicriteria Optimization.
\newblock 2nd ed. Springer Berlin, Heidelberg; 2005.
\newblock Originally published as volume 491 in the series: Lecture Notes in Economics and Mathematical Systems, XIII + 323 pages.
\newblock Available from: \url{https://doi.org/10.1007/3-540-27659-9}.

\bibitem{Harrigan2021}
Harrigan MP, Sung KJ, Neeley M, Satzinger KJ, Arute F, Arya K, et~al.
\newblock Quantum approximate optimization of non-planar graph problems on a planar superconducting processor.
\newblock Nature Physics. 2021 Mar;17(3):332-6.
\newblock Available from: \url{https://doi.org/10.1038/s41567-020-01105-y}.

\bibitem{Ebadi2022}
Ebadi S, Keesling A, Cain M, Wang TT, Levine H, Bluvstein D, et~al.
\newblock Quantum optimization of maximum independent set using Rydberg atom arrays.
\newblock Science. 2022;376(6598):1209-15.
\newblock Available from: \url{https://www.science.org/doi/abs/10.1126/science.abo6587}.

\bibitem{Zhu2026}
Zhu X, Zou Z, Jin F, Mosharev P, Luo M, Wu Y, et~al.
\newblock Combinatorial optimization enhanced by shallow quantum circuits with 104 superconducting qubits.
\newblock National Science Review. 2026 03:1-8.
\newblock Available from: \url{https://doi.org/10.1093/nsr/nwag124}.

\bibitem{Rajak2022}
Rajak A, Suzuki S, Dutta A, Chakrabarti BK.
\newblock Quantum annealing: an overview.
\newblock Philosophical Transactions of the Royal Society A: Mathematical, Physical and Engineering Sciences. 2022 12;381(2241):20210417.
\newblock Available from: \url{https://doi.org/10.1098/rsta.2021.0417}.

\bibitem{Kim2025}
Kim S, Ahn SW, Suh IS, Dowling AW, Lee E, Luo T.
\newblock Quantum annealing for combinatorial optimization: a benchmarking study.
\newblock npj Quantum Information. 2025 May;11(1):77.
\newblock Available from: \url{https://doi.org/10.1038/s41534-025-01020-1}.

\bibitem{Albash2018}
Albash T, Lidar DA.
\newblock Adiabatic quantum computation.
\newblock Rev Mod Phys. 2018 Jan;90.
\newblock Available from: \url{https://link.aps.org/doi/10.1103/RevModPhys.90.015002}.

\bibitem{Huang2023}
Huang T, Xu J, Luo T, Gu X, Goh R, Wong WF.
\newblock { Benchmarking Quantum(-Inspired) Annealing Hardware on Practical Use Cases}.
\newblock IEEE Transactions on Computers. 2023 Jun;72(06):1692-705.
\newblock Available from: \url{https://doi.ieeecomputersociety.org/10.1109/TC.2022.3219257}.

\bibitem{Zeng2024}
Zeng QG, Cui XP, Liu B, Wang Y, Mosharev P, Yung MH.
\newblock Performance of quantum annealing inspired algorithms for combinatorial optimization problems.
\newblock Communications Physics. 2024 Jul;7(1):249.
\newblock Available from: \url{https://doi.org/10.1038/s42005-024-01705-7}.

\bibitem{Mohseni2022}
Mohseni N, McMahon PL, Byrnes T.
\newblock Ising machines as hardware solvers of combinatorial optimization problems.
\newblock Nature Reviews Physics. 2022;4(6).
\newblock Available from: \url{https://par.nsf.gov/biblio/10402265}.

\bibitem{Kotil2025}
Kotil A, Pelofske E, Riedm{\"u}ller S, Egger DJ, Eidenbenz S, Koch T, et~al.
\newblock Quantum approximate multi-objective optimization.
\newblock Nature Computational Science. 2025 Dec;5(12):1168-77.
\newblock Available from: \url{https://doi.org/10.1038/s43588-025-00873-y}.

\bibitem{Farhi2014}
Farhi E, Goldstone J, Gutmann S.
\newblock A Quantum Approximate Optimization Algorithm.
\newblock arXiv preprint arXiv:14114028. 2014.

\bibitem{Blekos2024}
Blekos K, Brand D, Ceschini A, Chou CH, Li RH, Pandya K, et~al.
\newblock A review on Quantum Approximate Optimization Algorithm and its variants.
\newblock Physics Reports. 2024;1068:1-66.

\bibitem{king2025multi}
King AD.
\newblock Multi-objective optimization by quantum annealing.
\newblock arXiv preprint arXiv:251101762. 2025.

\bibitem{Kadowaki1998}
Kadowaki T, Nishimori H.
\newblock Quantum annealing in the transverse Ising model.
\newblock Phys Rev E. 1998 Nov;58:5355-63.
\newblock Available from: \url{https://link.aps.org/doi/10.1103/PhysRevE.58.5355}.

\bibitem{Johnson2011}
Johnson MW, Amin MHS, Gildert S, Lanting T, Hamze F, Dickson N, et~al.
\newblock Quantum annealing with manufactured spins.
\newblock Nature. 2011 May;473(7346):194-8.
\newblock Available from: \url{https://doi.org/10.1038/nature10012}.

\bibitem{goto2019}
Goto H, Tatsumura K, Dixon A.
\newblock Combinatorial optimization by simulating adiabatic bifurcations in nonlinear {Hamiltonian} systems.
\newblock Science Advances. 2019;5:1-8.
\newblock Available from: \url{https://www.semanticscholar.org/paper/dcf48728e86da649fc5367c219e76e54a596055a}.

\bibitem{goto2021}
Goto H, Endo K, Suzuki M, Sakai Y, Kanao T, Hamakawa Y, et~al.
\newblock High-performance combinatorial optimization based on classical mechanics.
\newblock Science Advances. 2021;7.
\newblock Available from: \url{https://www.semanticscholar.org/paper/bfe8e49dff8fd46b56678c0c531e9cb406af8ab5}.

\bibitem{BOLAND2017}
Boland N, Charkhgard H, Savelsbergh M.
\newblock A new method for optimizing a linear function over the efficient set of a multiobjective integer program.
\newblock European Journal of Operational Research. 2017;260(3):904-19.
\newblock Available from: \url{https://www.sciencedirect.com/science/article/pii/S0377221716300741}.

\bibitem{Dachert2024}
D{\"a}chert K, Fleuren T, Klamroth K.
\newblock A simple, efficient and versatile objective space algorithm for multiobjective integer programming.
\newblock Mathematical Methods of Operations Research. 2024 Aug;100(1):351-84.
\newblock Available from: \url{https://doi.org/10.1007/s00186-023-00841-0}.

\bibitem{zhang2007moea}
Zhang Q, Li H.
\newblock MOEA/D: A multiobjective evolutionary algorithm based on decomposition.
\newblock IEEE Transactions on evolutionary computation. 2007;11(6):712-31.

\bibitem{deb2002fast}
Deb K, Pratap A, Agarwal S, Meyarivan T.
\newblock A fast and elitist multiobjective genetic algorithm: NSGA-II.
\newblock IEEE transactions on evolutionary computation. 2002;6(2):182-97.

\bibitem{deb2013evolutionary}
Deb K, Jain H.
\newblock An evolutionary many-objective optimization algorithm using reference-point-based nondominated sorting approach, part I: solving problems with box constraints.
\newblock IEEE transactions on evolutionary computation. 2013;18(4):577-601.

\bibitem{jain2013evolutionary}
Jain H, Deb K.
\newblock An evolutionary many-objective optimization algorithm using reference-point based nondominated sorting approach, part II: Handling constraints and extending to an adaptive approach.
\newblock IEEE Transactions on evolutionary computation. 2013;18(4):602-22.

\bibitem{blank2019investigating}
Blank J, Deb K, Roy PC.
\newblock Investigating the normalization procedure of NSGA-III.
\newblock In: International conference on evolutionary multi-criterion optimization. Springer; 2019. p. 229-40.

\bibitem{cheng2016reference}
Cheng R, Jin Y, Olhofer M, Sendhoff B.
\newblock A reference vector guided evolutionary algorithm for many-objective optimization.
\newblock IEEE transactions on evolutionary computation. 2016;20(5):773-91.

\bibitem{Zitzler2003}
Zitzler E, Thiele L, Laumanns M, Fonseca CM, da~Fonseca VG.
\newblock Performance assessment of multiobjective optimizers: an analysis and review.
\newblock Trans Evol Comp. 2003 Apr;7(2):117–132.
\newblock Available from: \url{https://doi.org/10.1109/TEVC.2003.810758}.

\bibitem{Afsar2023}
Afsar B, Fieldsend JE, Guerreiro AP, Miettinen K, Rojas~Gonzalez S, Sato H.
\newblock In: Brockhoff D, Emmerich M, Naujoks B, Purshouse R, editors. Many-Objective Quality Measures. Cham: Springer International Publishing; 2023. p. 113-48.
\newblock Available from: \url{https://doi.org/10.1007/978-3-031-25263-1_5}.

\bibitem{Das_Dennis1998}
Das I, Dennis JE.
\newblock Normal-Boundary Intersection: A New Method for Generating the Pareto Surface in Nonlinear Multicriteria Optimization Problems.
\newblock SIAM Journal on Optimization. 1998;8(3):631-57.
\newblock Available from: \url{https://doi.org/10.1137/S1052623496307510}.

\bibitem{Blank2020}
Blank J, Deb K, Dhebar Y, Bandaru S, Seada H.
\newblock Generating Well-Spaced Points on a Unit Simplex for Evolutionary Many-Objective Optimization.
\newblock IEEE Transactions on Evolutionary Computation. 2020 01;PP:1-1.

\bibitem{moocore}
López-Ibáñez M, Fonseca CM, Paquete L, Guerreiro AP, Binois M, Bezerra LCT, et~al.. moocore; 2025.
\newblock Accessed 2026-04-15.
\newblock Available from: \url{https://github.com/multi-objective/moocore}.

\bibitem{tiunov2019a}
Tiunov E, Ulanov A, Lvovsky A.
\newblock Annealing by simulating the coherent Ising machine.
\newblock Optics Express. 2019;27(7):10288-95.

\bibitem{pymoo}
{Blank} J, {Deb} K.
\newblock pymoo: Multi-Objective Optimization in Python.
\newblock IEEE Access. 2020;8:89497-509.

\bibitem{DeSantis2026}
De~Santis M, L{\'e}tocart L, Zhang Y.
\newblock Quadratic convex reformulations for MultiObjective binary quadratic programming.
\newblock Journal of Global Optimization. 2026 Jan.
\newblock Available from: \url{https://doi.org/10.1007/s10898-025-01586-2}.

\bibitem{Quinton2025}
Quinton FA, Myhr PAS, Barani M, Crespo~del Granado P, Zhang H.
\newblock Quantum annealing applications, challenges and limitations for optimisation problems compared to classical solvers.
\newblock Scientific Reports. 2025 Apr;15(1):12733.
\newblock Available from: \url{https://doi.org/10.1038/s41598-025-96220-2}.

\bibitem{dwave_docs_advantage2_2025}
{D-Wave Systems Inc }. Performance gains in the D-Wave Advantage2 system at the 4,400-qubit scale; 2025.
\newblock Accessed 2026-04-10.
\newblock Available from: \url{https://www.dwavequantum.com/media/wakjcpsf/adv2_4400q_whitepaper-1.pdf}.

\bibitem{Kanao2022}
Kanao T, Goto H.
\newblock Simulated bifurcation assisted by thermal fluctuation.
\newblock Communications Physics. 2022 Jun;5(1):153.
\newblock Available from: \url{https://doi.org/10.1038/s42005-022-00929-9}.

\bibitem{Xiao2026}
Xiao Z, Huang Z, Qiu Q, Liu YQ, Zhuang JP, Yung MH.
\newblock Globally guided simulated bifurcation for enhanced optimization.
\newblock Phys Rev Appl. 2026 Feb;25:024014.
\newblock Available from: \url{https://link.aps.org/doi/10.1103/sdb2-stbs}.

\bibitem{Tao2026}
Tao XZ, Zeng QG, Huang ZJ, Zuo BW, Liu YQ, Zhuang J, et~al.
\newblock Tabu-Enhanced Simulated Bifurcation for combinatorial optimization.
\newblock Communications Physics. 2026 Feb;9(1):100.
\newblock Available from: \url{https://doi.org/10.1038/s42005-026-02538-2}.

\bibitem{SCIPOptSuite10}
Hojny C, Besan{\c{c}}on M, Bestuzheva K, Borst S, Chmiela A, Dion{\'{i}}sio J, et~al.
\newblock {The SCIP Optimization Suite 10.0}.
\newblock Optimization Online; 2025.
\newblock Available from: \url{https://optimization-online.org/2025/11/the-scip-optimization-suite-10-0/}.

\bibitem{xu2024mindspore}
Xu X, Cui J, Cui Z, He R, Li Q, Li X, et~al.
\newblock MindSpore Quantum: a user-friendly, high-performance, and AI-compatible quantum computing framework.
\newblock arXiv preprint arXiv:240617248. 2024.

\end{thebibliography}

\clearpage

\onecolumn
\appendix

\section{Hypervolume indicator}
\label{app:Hypervolume}

The hypervolume (HV) indicator is a widely used scalar measure for comparing
finite approximation fronts in multi-objective optimization. In this work, HV is
computed in the original cut-value maximization space. Let
$\widehat{\mathcal F}\subset\mathbb R^K$ denote the set of nondominated
objective vectors returned by the algorithm, and let
$\mathbf r\in\mathbb R^K$ be a reference point dominated by all vectors in
$\widehat{\mathcal F}$, i.e., $r_k\le y_k$ for all
$\mathbf y\in\widehat{\mathcal F}$ and all $k$. The hypervolume is the
$K$-dimensional Lebesgue measure of the region dominated by
$\widehat{\mathcal F}$ and bounded by $\mathbf r$:
\[
HV(\widehat{\mathcal F},\mathbf r)
=
\Lambda\left(
\bigcup_{\mathbf y\in\widehat{\mathcal F}}
[\mathbf r,\mathbf y]
\right),
\qquad
[\mathbf r,\mathbf y]
=
\prod_{k=1}^{K}[r_k,y_k].
\]
For a fixed reference point, HV is Pareto-compliant: if one approximation front
dominates another, then it has a larger hypervolume. The indicator therefore
combines information about convergence and spread of the approximation front,
although its numerical value depends on the choice of reference point.

For illustration, consider a biobjective maximization problem with reference
point $\mathbf r=(0,0)$ and two nondominated objective vectors
$\mathbf y_A=(10,5)$ and $\mathbf y_B=(5,10)$. The two dominated rectangles have
areas $50$ and $50$, and their overlap has area $25$. Hence,
\[
HV(\{\mathbf y_A,\mathbf y_B\},\mathbf r)
=
50+50-25
=
75.
\]
Adding a third nondominated vector $\mathbf y_C=(8,8)$ increases the dominated
area to $84$, since it contributes the additional region
$[5,8]\times[5,8]$.

\section{Das-Dennis simplex lattice and exact Pareto-optimal solutions}
\label{app:das-dennis}

\begin{figure*}[h]
    \centering
    \includegraphics[width=1\textwidth]{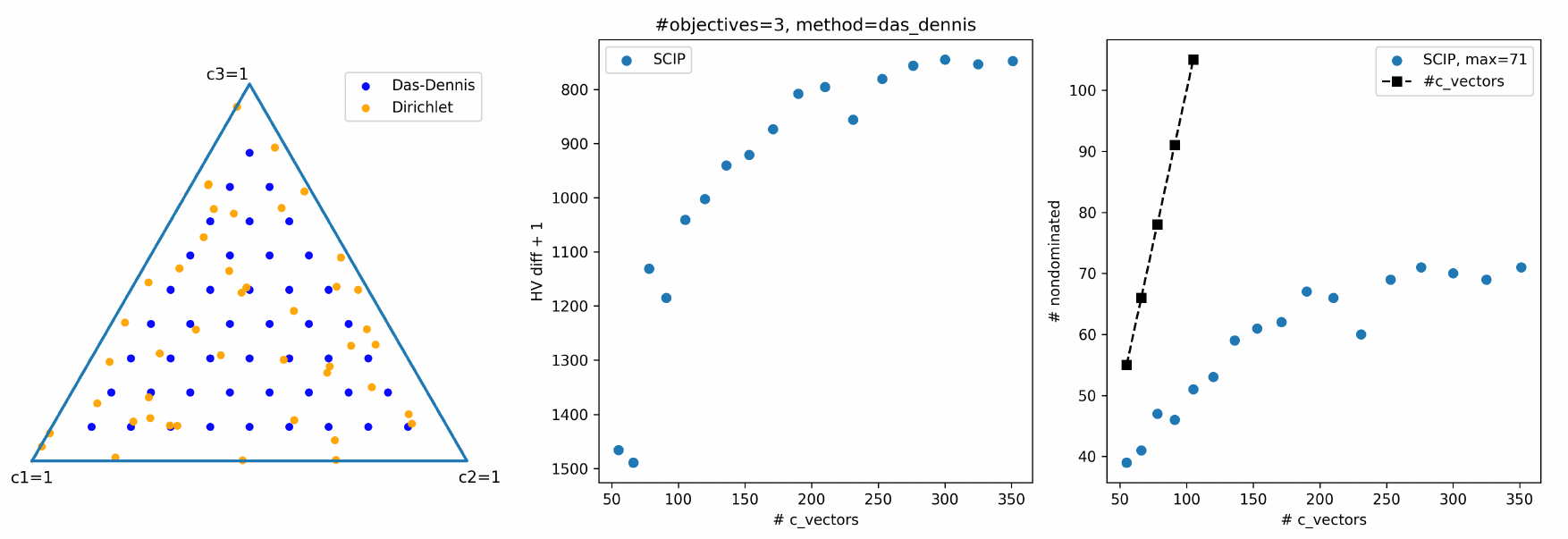}
    \caption{\textbf{Left:} Das-Dennis simplex lattice points compared with samples from a random Dirichlet distribution for three-objective problem. \textbf{Middle:} Hypervolume difference for solutions found by the exact solver as a function of the number of fixed weight vectors. \textbf{Right:} Number of Pareto-optimal solutions found by exact solver at each number of fixed weight vectors.}
    \label{fig:Simplex_lattice}
\end{figure*}

In Fig.~\ref{fig:Simplex_lattice} we show an example of a simplex lattice used to produce weight vectors for scalarizing the three-objective problem. We also solved corresponding single-objective problems to optimality using the exact MIP/MINLP solver SCIP \cite{SCIPOptSuite10}, and show that the number of found Pareto-optimal solutions and the corresponding hypervolume grow much slower than the number of vectors in the lattice. This indicates that the majority of Pareto-optimal solutions do not correspond to an exact solution of the scalarized problem with any {weight vector in the tested lattice}. Thus an approximate algorithm such as NI-SB that samples a wide variety of near-optimal solutions is well suited for exploration of the Pareto set.

% \clearpage
\newpage

\section{Noise amplitude and effective sample size}
\label{app:noise}
\begin{figure*}[ht]
    \centering
    \includegraphics[width=1\textwidth]{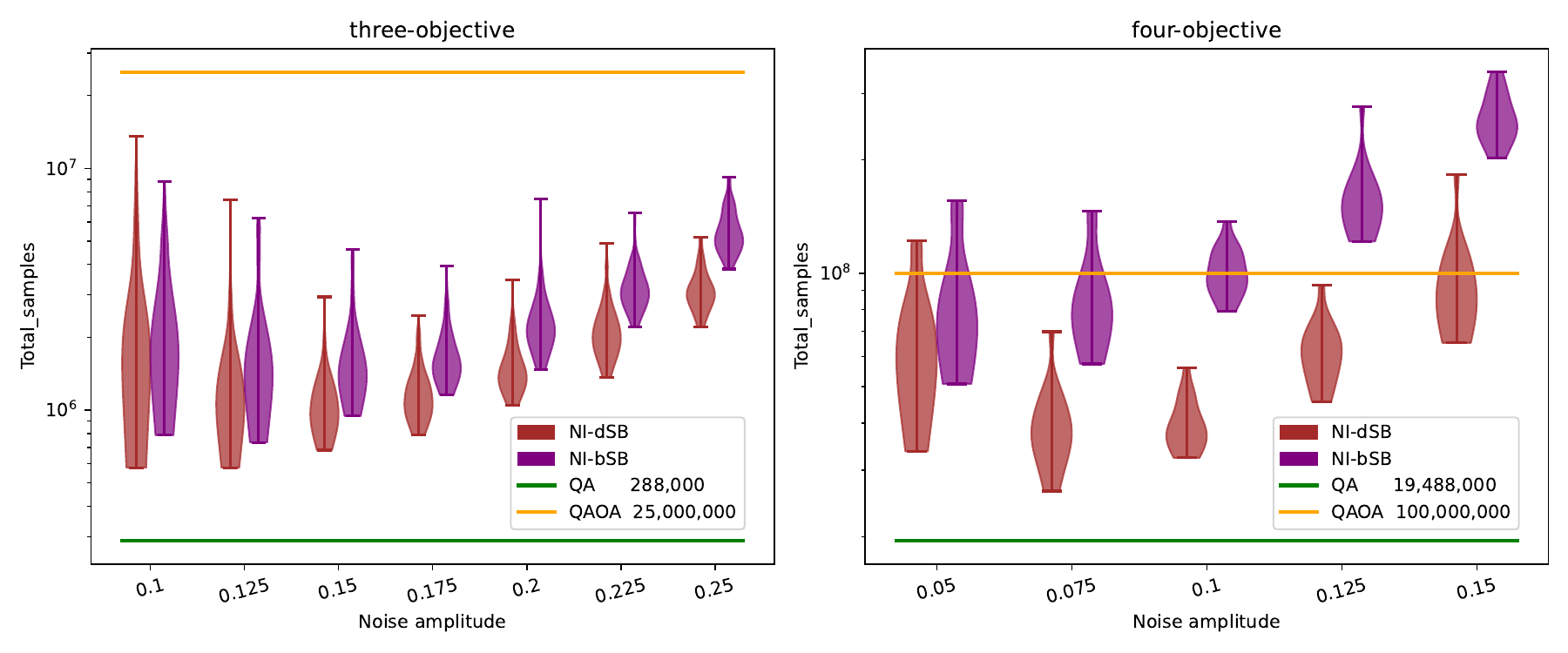}
    \caption{Total sample size required to reach {the best-known} hypervolume, as a function of the noise amplitude injected in the algorithm. The QA and QAOA values are taken from Refs.~\cite{king2025multi, Kotil2025}.}
    \label{fig:sample_quality_by_noise_amp}
\end{figure*}

In Fig.~\ref{fig:sample_quality_by_noise_amp} we illustrate the choice of optimal noise amplitude $\alpha$ in Eq.~(\ref{eq:NI-SB_equation}) for both NI-bSB and NI-dSB. The figure shows that the main results of this paper (fast generation of samples that contain Pareto-optimal set, with the number of samples required to reproduce the whole Pareto front lying between the reported QA and QAOA results) hold over a wide range of noise amplitudes for the three-objective problem and are achievable with a {noise amplitude chosen by the following criterion} for the four-objective problem. {For each value of $\alpha$ we measured the total number of samples needed to reach the reported Pareto front, and selected the $\alpha$ that minimizes this quantity.} The value of $\alpha$ {that minimizes the samples-to-target criterion lies} between $0.125$ and $0.2$ for both bSB and dSB in three-objective case, and around $0.1$ for four-objective{; the broad plateau in Fig.~\ref{fig:sample_quality_by_noise_amp} indicates that the conclusions do not depend sensitively on this choice}.

\section{Comparison with Simulated Coherent Ising Machine (SimCIM) and Simulated Annealing (SA)}
\label{app:simcim}
Simulated Coherent Ising Machine (SimCIM) is a quantum-inspired optimization algorithm~\cite{tiunov2019a} that shares similarities with noise-injected simulated bifurcation methods. Both algorithms feature iterative update rules and incorporate noise terms in their formulations.
We evaluate SimCIM and NI-SB on the three-objective problem from~\cite{Kotil2025} using identical hardware. The results demonstrate that NI-SB achieves faster convergence than SimCIM. The key distinction lies in how momentum is introduced: in SB algorithms, momentum naturally emerges from the Hamiltonian equations and is set to zero when spin variables exceed the [-1,1] range, whereas in SimCIM momentum results from applying momentum gradient descent to the state-update equation without such boundary constraints. Our implementation of SimCIM utilizes MindSpore Quantum~\cite{xu2024mindspore}.

We also add results for Simulated Annealing, another common metaheuristic widely adopted for finding or approximating ground states of Ising Hamiltonians in the single-objective case. Execution of SA is inherently sequential, which makes GPU acceleration difficult, unlike SB and SimCIM. Thus we present results for SA executed on CPU (laptop AMD Ryzen AI 7 PRO 350), and add for reference results of NI-SB executed on the same machine. We used SA as implemented in D-Wave's Neal library, set the number of sweeps and the batch size to values similar to those used for SB, and inverse temperature range $[0.1, 2]$.

We selected relatively high temperatures for SA by hand, and set the noise amplitude in SimCIM to the same value with SB. This experiment demonstrates why we choose SB as the primary candidate for our work. Comprehensive comparison, including other variants of QAIA may be the subject of future work.

\begin{figure*}[ht]
    \centering
    \includegraphics[width=0.5\textwidth]{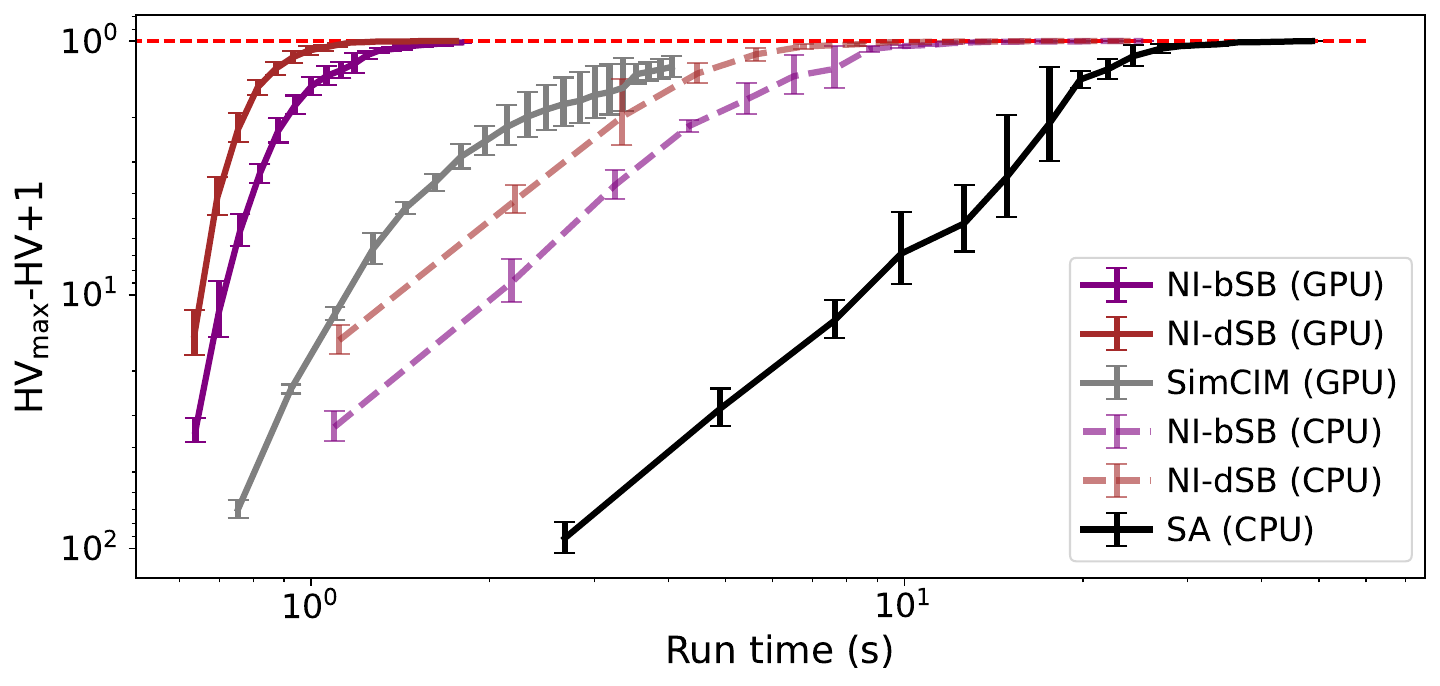}
    \caption{Comparison of SA, SimCIM and NI-SB. The y-axis shows the Hypervolume difference ($HV_{\text{max}} - HV_t + 1$) on a logarithmic scale, and the x-axis represents the end-to-end run time in seconds.}
    \label{fig:simcim}
\end{figure*}

\section{Solution Distribution and Pareto Front of MO-MaxCut Instances}
\label{app:3D_front}

\begin{figure*}[ht]
    \centering
    \includegraphics[width=0.5\textwidth]{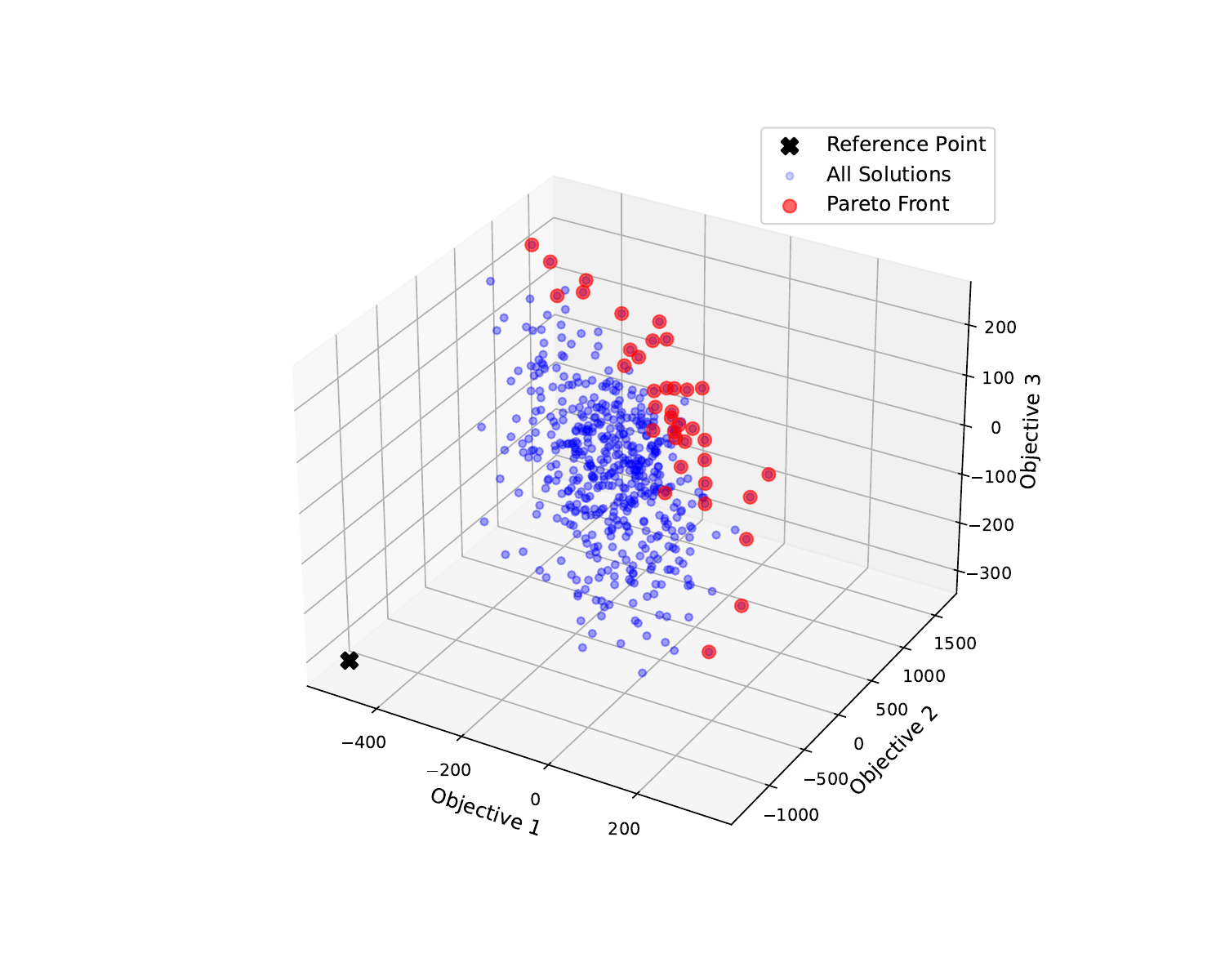}
    \caption{Solution space of a random MO-MaxCut instance with $n=10$ variables. The blue points represent all feasible solutions, while the red points highlight the Pareto front.}
    \label{fig:3D_front}
\end{figure*}

The MO-MaxCut instances for the experiment in Table~\ref{tab:MOEA_compare} are generated randomly. Specifically, we first generate two $n \times n$ matrices, $A$ and $B$, with integer elements uniformly sampled from the interval $[-25, 25]$. We then construct two new matrices, $J^{(1)}$ and $J^{(2)}$, via a linear transformation:
\begin{equation}
    \begin{pmatrix} J^{(1)} \\ J^{(2)} \end{pmatrix} 
    = 
    \begin{pmatrix} a & b \\ c & d \end{pmatrix} 
    \begin{pmatrix} A \\ B \end{pmatrix},
\end{equation}
where the parameters are set to $a=1, b=1, c=0.2, d=-5$. This configuration results in a negative correlation between the cut values $C_1$ and $C_2$, with a Pearson correlation coefficient of approximately $-0.7$. Here, $C_i$ is defined by Eq.~\ref{eq:single_obj}. The third objective matrix, $J^{(3)}$, is generated independently using the same method as $A$ and $B$. Additionally, a random mask matrix is applied to control the edge density of the graph. Figure~\ref{fig:3D_front} illustrates the solution distribution for a random MO-MaxCut instance with $n=10$ variables. 

\section{Standard deviations for Table~\ref{tab:MOEA_compare}}
\label{app:table2_std}

Table~\ref{tab:MOEA_std} reports the standard deviations corresponding to the means in Table~\ref{tab:MOEA_compare}. All entries are sample standard deviations over the five independent runs described in the main text. For the $n=25$ and $n=42$ instances, the reference front is known and every run of NI-bSB and NI-dSB reaches the same hypervolume, so the standard deviations of their hypervolume ratios are zero. On the composite rows, the standard deviations quantify the run-to-run variability of both the achieved hypervolume and the number of recovered archive points.

\begin{table}
\resizebox{\linewidth}{!}{
\begin{tabular}{cc|cc|cc|cc|cc|cc|cc|cc}
\hline
\multicolumn{2}{c|}{Instance} & \multicolumn{2}{c|}{Random} & \multicolumn{2}{c|}{RVEA} & \multicolumn{2}{c|}{MOEA/D} & \multicolumn{2}{c|}{NSGAII} & \multicolumn{2}{c|}{NSGAIII} & \multicolumn{2}{c|}{NI-bSB} & \multicolumn{2}{c}{NI-dSB} \\ \hline
$n$ & $d$ & \begin{tabular}[c]{@{}c@{}}Front \\ points\end{tabular} & \begin{tabular}[c]{@{}c@{}}HV\\ Ratio\end{tabular} & \begin{tabular}[c]{@{}c@{}}Front \\ points\end{tabular} & \begin{tabular}[c]{@{}c@{}}HV\\ Ratio\end{tabular} & \begin{tabular}[c]{@{}c@{}}Front \\ points\end{tabular} & \begin{tabular}[c]{@{}c@{}}HV\\ Ratio\end{tabular} & \begin{tabular}[c]{@{}c@{}}Front \\ points\end{tabular} & \begin{tabular}[c]{@{}c@{}}HV\\ Ratio\end{tabular} & \begin{tabular}[c]{@{}c@{}}Front \\ points\end{tabular} & \begin{tabular}[c]{@{}c@{}}HV\\ Ratio\end{tabular} & \begin{tabular}[c]{@{}c@{}}Front \\ points\end{tabular} & \begin{tabular}[c]{@{}c@{}}HV\\ Ratio\end{tabular} & \begin{tabular}[c]{@{}c@{}}Front \\ points\end{tabular} & \begin{tabular}[c]{@{}c@{}}HV\\ Ratio\end{tabular} \\ \hline
25 & 0.5 & $\pm$4.8 & $\pm$0.7 & $\pm$2.8 & $\pm$1.1 & $\pm$6.9 & $\pm$2.8 & $\pm$7.3 & $\pm$0.6 & $\pm$3.4 & $\pm$0.3 & $\pm$0.9 & $\pm$0.0 & $\pm$0.0 & $\pm$0.0 \\
25 & 1.0 & $\pm$2.8 & $\pm$0.9 & $\pm$3.4 & $\pm$1.4 & $\pm$5.5 & $\pm$3.9 & $\pm$7.6 & $\pm$1.4 & $\pm$3.9 & $\pm$0.3 & $\pm$0.0 & $\pm$0.0 & $\pm$0.0 & $\pm$0.0 \\ \hline
42 & \begin{tabular}[c]{@{}c@{}}heavy-hex\end{tabular} & $\pm$0.0 & $\pm$0.8 & $\pm$7.9 & $\pm$0.9 & $\pm$3.6 & $\pm$1.3 & $\pm$15.7 & $\pm$1.3 & $\pm$10.9 & $\pm$0.6 & $\pm$0.0 & $\pm$0.0 & $\pm$0.0 & $\pm$0.0 \\ \hline
100 & 0.5 & $\pm$0.0 & $\pm$0.3 & $\pm$0.0 & $\pm$6.0 & $\pm$0.0 & $\pm$3.1 & $\pm$9.0 & $\pm$4.9 & $\pm$16.0 & $\pm$3.9 & $\pm$40.1 & $\pm$0.1 & $\pm$64.0 & $\pm$0.1 \\
100 & 1.0 & $\pm$0.0 & $\pm$0.2 & $\pm$1.8 & $\pm$6.3 & $\pm$0.9 & $\pm$2.3 & $\pm$3.5 & $\pm$4.9 & $\pm$21.6 & $\pm$6.1 & $\pm$51.6 & $\pm$0.3 & $\pm$62.7 & $\pm$0.1 \\
200 & 0.5 & $\pm$0.0 & $\pm$0.4 & $\pm$6.0 & $\pm$3.8 & $\pm$0.0 & $\pm$3.0 & $\pm$17.3 & $\pm$5.8 & $\pm$14.8 & $\pm$3.4 & $\pm$37.0 & $\pm$0.5 & $\pm$94.1 & $\pm$0.6 \\
200 & 1.0 & $\pm$0.0 & $\pm$0.8 & $\pm$15.4 & $\pm$5.5 & $\pm$3.9 & $\pm$5.0 & $\pm$26.7 & $\pm$10.6 & $\pm$16.9 & $\pm$3.8 & $\pm$31.7 & $\pm$2.2 & $\pm$76.9 & $\pm$2.9 \\ \hline
\end{tabular}
}
\caption{Standard deviations of the quantities reported in Table~\ref{tab:MOEA_compare}, computed over five independent runs. Each entry is the standard deviation of the corresponding mean, preceded by $\pm$, in numbers of solutions for the front-point columns and in percentage points for the hypervolume-ratio columns.}
\label{tab:MOEA_std}
\end{table}

\end{document}